\documentclass[pra, twocolumn, nofootinbib, amsfonts, superscriptaddress]{revtex4}

\usepackage[dvips]{graphicx}
\usepackage{amsmath}
\usepackage{amsopn}

\newtheorem{theorem}{Theorem}

\newtheorem{proposition}[theorem]{Proposition}
\newcommand{\qed}{\rule{7pt}{7pt}}

\def\<{\langle}
\def\>{\rangle}
\def\cO{{\mathcal O}}
\def\be{\begin{equation}}
\def\ee{\end{equation}}

\newcommand{\bra}[1]{\mbox{$\langle #1 |$}}
\newcommand{\ket}[1]{\mbox{$| #1 \rangle$}}
\newcommand{\braket}[2]{\mbox{$\langle #1 | #2 \rangle$}}
\newcommand{\eq}[1]{Eq.~(\ref{eq:#1})}
\newcommand{\fig}[1]{Fig.~(\ref{fig:#1})}

\DeclareMathOperator{\tr}{tr}

\begin{document}

\title{Universal quantum data compression via gentle tomography}
\author{Charles H. Bennett}
\email{bennetc@watson.ibm.com}
\affiliation{IBM T.J. Watson Research Center, P.O. Box 218, Yorktown
Heights, NY 10598, USA}
\author{Aram W. Harrow}
\email{aram@mit.edu}
\affiliation{MIT Physics Dept., 77 Massachusetts Ave, Cambridge, MA
02139, USA}
\author{Seth Lloyd}
\email{slloyd@mit.edu}
\affiliation{MIT Dept of Mechanical Engineering}
\date{\today}

\begin{abstract}
Quantum state tomography--the practice of estimating a quantum state
by performing measurements on it--is useful in a variety of contexts.
We introduce ``gentle tomography'' as a version of
tomography that preserves the measured quantum data.  As an
application of gentle tomography, we describe a polynomial-time method
for universal source coding. 
\end{abstract}
\maketitle

\section{Introduction}
Suppose that we have a sequence of quantum states, each drawn from an
ensemble with known density matrix $\rho$.  Schumacher compression
then allows the sequence to be efficiently encoded so that $S(\rho) =
-\tr \rho \log_2 \rho$ qubits are required to encode each state in the
limit that the length of the sequence goes to infinity \cite{S95}.
This resembles classical source coding, in which a source can be
compressed to a rate asymptotically approaching its Shannon entropy.
However, classical compression can be performed by algorithms that are
{\em universal} (do not depend on a description of the source) and
{\em efficient} (have running time polynomial in the length of the
input).  In contrast, most existing quantum compression algorithms
either rely on knowing the basis in which $\rho$ is diagonal
\cite{S95} or have no known polynomial time implementations
\cite{JHHH98, HM02a}. 

This paper presents an efficient, universal, quantum data compression
algorithm; that is, it can compress an unknown i.i.d. quantum source
$\rho^{\otimes n}$ in $\text{poly}(n)$ time to a rate converging to
its von Neumann entropy $S(\rho)$ and with error approaching zero as
the number of copies, $n$, increases.  Another efficient universal
quantum data compression algorithm was presented in \cite{JP02}, but
our algorithm has the advantages of simplicity and a better
rate-disturbance tradeoff.

Our algorithm consists of two parts: a weak measurement of
$\rho^{\otimes n}$ that estimates $\rho$ accurately without causing
very much damage to the state, followed by compressing $\rho^{\otimes
n}$ based on this estimate.  Conceptually, this resembles classical
methods of compression which determine the empirical distribution
of their input in their first pass over the data and perform the
compression in the second pass.  The only new difficulties we will
encounter in the quantum case come from the need to perform state
tomography on $\rho$ without causing very much damage and to compress
$\rho$ based on an imperfect estimate.

\section{Gentle tomography}

The problem of weakly measuring states of the form $\rho^{\otimes n}$
was introduced in \cite{CTDSZ80} and further developed in
\cite{LS00,HM02a}.  While it is impossible to measure a single state
$\rho$ without causing disturbance, we expect ordinary classical logic
to apply to $\rho^{\otimes n}$ when $n$ is large, so that it is
possible to measure even non-commuting observables precisely with
little disturbance.  For example, in Nuclear Magnetic Resonance, the
total $x$-magnetization of $n=\cO(10^{20})$ nuclear spins is
continuously measured without causing decoherence by a probe
consisting of a coil of wire around the sample.  This is possible
because the measurement does not precisely determine the number of
nuclear spins pointing in the $x$ direction, but only gives a crude
estimate of the quantity.  In this section, we will introduce a
procedure for state tomography on $\rho^{\otimes n}$ and then show how
to modify it so its disturbance vanishes for large $n$ while at the
same time it yields an asymptotically accurate estimate of $\rho$.

Let $\{\sigma_k\}_{k=1}^{d^2-1}$ is an orthonormal
($\tr\sigma_j\sigma_k=\delta_{jk}$) basis of traceless Hermitian
$d\times d$ matrices, and write the density matrix $\rho$ as
$\rho=I/d+\sum_k (\tr \rho\sigma_k)\sigma_k$.  Estimating $\rho$
reduces to estimating the $d^2-1$ quantities $\tr\rho\sigma_k$.  If we
now diagonalize $\sigma_k$ as $\sigma_k=\sum_{i=1}^d \lambda_i
\ket{v_i}\bra{v_i}$, then $\tr\rho\sigma_k=\sum_i \lambda_i
\bra{v_i}\rho\ket{v_i}$, so state tomography reduces to estimating
$d(d^2-1)$ quantites of the form $\bra{\phi}\rho\ket{\phi}$ and then
performing a classical computation.\footnote{Modifying our techniques
to only estimate the $d^2-1$ quantities $\tr\rho\sigma_k$ would cause
the state estimate to converge more quickly, but this would make our
exposition slightly more complicated.  Unfortunately, there is no
known polynomial time implementation of quantum state tomography that
has the probability of large deviations vanish at the asymptotically
optimal rate.}


If we didn't mind damaging the state, then one method of estimating
$\alpha:=\bra{\phi}\rho\ket{\phi}$ would be to apply the projective
measurement $\{\ket{\phi}\bra{\phi}, I - \ket{\phi}\bra{\phi}\}$ to
each copy of $\rho$.  The number of occurences of
$\ket{\phi}\bra{\phi}$ would be binomially distributed with mean
$n\alpha$ and variance $n\alpha(1-\alpha)\leq n/4$, so we could
reliably estimate $\alpha$ to an accuracy of $\cO(n^{-1/2})$.  Of
course, this measurement would drastically damage some states, such as $\frac{1}{\sqrt{2}}(\ket{\phi}+\ket{\phi^\perp})$.

\begin{figure}[tbph]
\setlength{\unitlength}{3947sp}%
\begingroup\makeatletter\ifx\SetFigFont\undefined%
\gdef\SetFigFont#1#2#3#4#5{%
  \reset@font\fontsize{#1}{#2pt}%
  \fontfamily{#3}\fontseries{#4}\fontshape{#5}%
  \selectfont}%
\fi\endgroup%
\begin{picture}(6837,1295)(76,-673)
\thinlines
\put(1437,281){\circle*{40}}
\put(1576,189){\circle*{40}}
\put(1726,94){\circle*{40}}
\put(2776,-661){\oval(300,300)[tr]}
\put(2776,-661){\oval(300,300)[tl]}
\put(2626,-661){\framebox(300,300){}}
\put(2761,-661){\vector( 1, 2){128}}
\put(451,277){\oval(152,524)[tl]}
\put(451,277){\oval(152,526)[bl]}
\put(826,539){\circle*{150}}
\put(1276,389){\circle*{150}}
\put(2123, -8){\circle*{150}}
\put(526,539){\line( 1, 0){2625}}
\put(526,389){\line( 1, 0){2625}}
\put(601,-511){\line( 1, 0){ 75}}
\put(976,-511){\line( 1, 0){150}}
\put(1426,-511){\line( 1, 0){525}}
\put(2251,-511){\line( 1, 0){300}}
\put(526, 14){\line( 1, 0){2625}}
\put(826,-361){\line( 0, 1){900}}
\put(676,-661){\framebox(300,300){}}
\put(1276,-361){\line( 0, 1){750}}
\put(1126,-661){\framebox(300,300){}}
\put(2101,-361){\line( 0, 1){375}}
\put(1951,-661){\framebox(300,300){}}
\put( 200,314){\makebox(0,0)[c]{$\rho^{\otimes n}$}}
\put( 376,-586){\makebox(0,0)[lb]{$|0\rangle$}}
\put(706,-586){\makebox(0,0)[lb]{+1}}
\put(1156,-586){\makebox(0,0)[lb]{+1}}
\put(1981,-601){\makebox(0,0)[lb]{+1}}
\end{picture}
\caption{Circuit for performing the measurement in
\eq{collective-measurement}.  The controlled-$+1$ operations map
$\ket{\phi}\ket{x}$ to $\ket{\phi}\ket{x+1}$ for any value $x$ of the
target and leave other states unchanged.}
\label{fig:measure}
\end{figure}
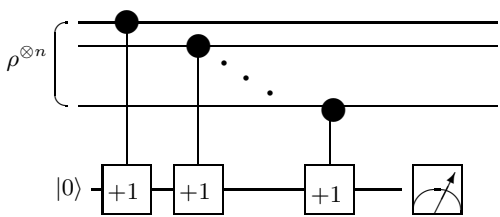

Instead of measuring each state individually, we can also express this
measurement as a collective operation on all $n$ states
simultaneously.  It is given by the operators
\be
M_k=\sum_{\substack{x\in\{0,1\}^n\\|x|=k}}
\bigotimes_{i=1}^n x_i\ket{\phi}\bra{\phi} + 
(1-x_i)(\openone - \ket{\phi}\bra{\phi}).
\label{eq:collective-measurement}\ee
where $k$ ranges from $0$ to $n$ and $|x|$ denotes the number of 1's
in the $n$-bit string $x$.  Clearly, measuring $\{M_k\}$ yields the same
statistics as measuring each state individually and counting the
$\ket{\phi}\bra{\phi}$ outcomes.  The measurement can also be
constructed efficiently: we unitarily count the number of occurences
of $\ket{\phi}$ in the $n$ states in an ancilla register and then
measure the ancilla (see Fig.~\ref{fig:measure}).

Unfortunately, even the collective measurement in
\eq{collective-measurement} causes substantial damage to the state.
For example, if the measurement $\{M_k\}$ is repeated, then the
distribution of $k$ will have a variance
of $O(n)$ the first time and $0$ on subsequent
measurements.\footnote{This damage is sometimes useful.  In
\cite{BBPS95}, it is used as the first step of entanglement
concentration.  In fact, our compression protocol may be thought of as
the gentle analogue of \cite{BBPS95} in the same way that the
compression scheme in \cite{HM02a} is the gentle analogue of the
entanglement concentration procedure of \cite{HM02b}.}

In \cite{LS00} this problem was solved by initalizing the ancilla in
\fig{measure} to the state $\sum_k e^{-k^2/2\Delta^2}\ket{k}$ instead
of $\ket{0}$.  The measurement of $k$ then has variance
$\Delta^2+O(n)$ and it can be shown\cite{HL04} that the damage to
$\rho^{\otimes n}$ is $O(n/(\Delta^2+n))$. Ref.~\cite{HM02a} proposed
a method which causes more damage to the state, but is easier to
analyze for our purposes.

To implement the gentle measurement of \cite{HM02a}, we will divide up
the range from $0\ldots n$ into $m$ bins, with boundaries $0=b_0 \leq
b_1 \cdots \leq b_m=n+1$.  Then we will modify the collective
measurement of \eq{collective-measurement} to measure only the bin
that the state lies in instead of determining the exact value of $k$.
The new measurement $\{M'_j\}$ is given in terms of the $M_k$ of
\eq{collective-measurement} by
\be M'_j = \sum_{b_{j-1}\leq k < b_j} M_k 
\label{eq:gentle-measurement}\ee
where $j$ ranges from 1 to $m$.

If the bin size, $n/m$, is much larger than the
$\cO(\sqrt{n})$ width of $\rho^{\otimes n}$ then we expect to project
onto a measurement outcome that contains almost all of the support of
$\rho^{\otimes n}$, thereby causing little disturbance.  Since we want
to avoid having a bin boundary within $\cO(\sqrt{n})$ of the state,
{\em for any choice of $\rho$}, we will choose the $b_i$ uniformly at
random from between 0 and $n$.  

The choice of $m$ now defines a trade-off between disturbance caused
to $\rho^{\otimes n}$ and information gained about $\rho$.  Choosing
a smaller $m$ means that each bin is larger, so that a measurement
outcome lets us infer less about $\rho$, but we have a smaller
probability of damaging $\rho^{\otimes n}$ by projecting onto only
part of its support.

\begin{proposition}\label{prop:gentle}
The measurement $\{M_j'\}$ described above can be implemented in
$\cO(n)$ gates.  If we choose $m=n^s$ for $0<s<1/2$, then the
measurement will fail with probability $\cO(n^{s-1/2}\ln n)$.  Upon success, the
measurement outcome is within $\cO(n^{1-s}\ln n)$ of $n\alpha$ and the
disturbance (in the sense of entanglement fidelity) is less than
$\exp(-\cO(\ln^2n))\leq \cO(n^{-p})$ for any constant $p$.
\end{proposition}

{\bf Proof of proposition~\ref{prop:gentle}:} 

\begin{figure}[htbp]
\setlength{\unitlength}{3000sp}%
\begingroup\makeatletter\ifx\SetFigFont\undefined%
\gdef\SetFigFont#1#2#3#4#5{%
  \reset@font\fontsize{#1}{#2pt}%
  \fontfamily{#3}\fontseries{#4}\fontshape{#5}%
  \selectfont}%
\fi\endgroup%
\begin{picture}(6837,1670)(76,-1048)
\thinlines
\put(3751,-1036){\oval(300,300)[tr]}
\put(3751,-1036){\oval(300,300)[tl]}
\put(3601,-1036){\framebox(300,300){}}
\put(3736,-1036){\vector( 1, 2){128}}
\put(1457,301){\circle*{40}}
\put(1596,209){\circle*{40}}
\put(1746,134){\circle*{40}}
\put(4232,301){\circle*{40}}
\put(4371,209){\circle*{40}}
\put(4521,134){\circle*{40}}
\put(451,277){\oval(152,524)[tl]}
\put(451,277){\oval(152,526)[bl]}
\put(3601,539){\circle*{150}}
\put(4051,389){\circle*{150}}
\put(4748, -8){\circle*{150}}
\put(826,539){\circle*{150}}
\put(1276,389){\circle*{150}}
\put(2123, -8){\circle*{150}}
\put(526,539){\line( 1, 0){4425}}
\put(526,389){\line( 1, 0){4425}}
\put(601,-511){\line( 1, 0){ 75}}
\put(976,-511){\line( 1, 0){150}}
\put(1426,-511){\line( 1, 0){525}}
\put(2251,-511){\line( 1, 0){300}}
\put(526, 14){\line( 1, 0){4425}}
\put(601,-886){\line( 1, 0){1950}}
\put(2551,-1036){\framebox(600,750){}}
\put(3151,-886){\line( 1, 0){450}}
\put(3151,-511){\line( 1, 0){300}}
\put(3751,-511){\line( 1, 0){150}}
\put(4201,-511){\line( 1, 0){375}}
\put(3601,-361){\line( 0, 1){900}}
\put(4051,-361){\line( 0, 1){750}}
\put(4726,-361){\line( 0, 1){375}}
\put(3451,-661){\framebox(300,300){}}
\put(3901,-661){\framebox(300,300){}}
\put(4576,-661){\framebox(300,300){}}
\put(826,-361){\line( 0, 1){900}}
\put(676,-661){\framebox(300,300){}}
\put(1276,-361){\line( 0, 1){750}}
\put(1126,-661){\framebox(300,300){}}
\put(2101,-361){\line( 0, 1){375}}
\put(1951,-661){\framebox(300,300){}}
\put( 10,224){\makebox(0,0)[lb]{$\rho^{\otimes n}$}}
\put( 76,-586){\makebox(0,0)[lb]{$|0\rangle$}}
\put( 76,-961){\makebox(0,0)[lb]{$|0\rangle$}}
\put(2626,-736){\makebox(0,0)[lb]{{\sc BIN}}}
\put(706,-586){\makebox(0,0)[lb]{+1}}
\put(1156,-586){\makebox(0,0)[lb]{+1}}
\put(1981,-601){\makebox(0,0)[lb]{+1}}
\put(3496,-601){\makebox(0,0)[lb]{-1}}
\put(3946,-601){\makebox(0,0)[lb]{-1}}
\put(4636,-601){\makebox(0,0)[lb]{-1}}
\end{picture}
\caption{Circuit for performing the gentle measurement in
\eq{gentle-measurement}.  The controlled-$+1$ and controlled-$-1$
operations act on the target only when the control is in the state
$\ket{\phi}$.  The gate {\sc BIN} classically computes which bin
contains the top register and stores it in the bottom register.}
\label{fig:gentle-measure}
\end{figure}
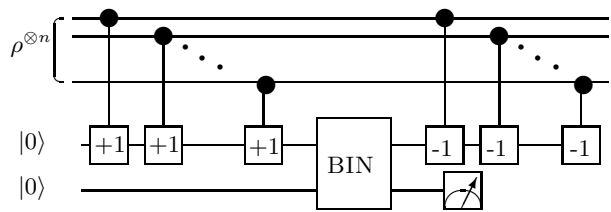

We begin by describing how to implement $\{M_j'\}$.  First we count
the number of times $\ket{\phi}$ occurs in $\rho^{\otimes n}$ and
store the result $k\in\{0,\ldots,n\}$ in an ancilla register.  Then we perform
a classical computation to determine which bin $j$ contains the
result $k$.  We measure $j$, thus implementing the projective
measurement $M'_j$ and then uncompute $j$ and finally uncompute $k$.
This is demonstrated in Fig~\ref{fig:gentle-measure}

We define three possible causes of failure: i) some $b_i$ will be too
close to $n\alpha$ (within $n^{1/2}\ln n$), ii) there won't be any
$b_i$ on either side of $n\alpha$ within $n^{1-s}\ln n$ and iii)
measuring $M_j'$ will yield a bin that does not contain $n\alpha$.
Using the union bound, we can show that i) has probability $\leq
m(2n^{1/2}\ln n)/n = 2n^{s-1/2}\ln n$.  Next, the probability that no
$b_i$ is in $[n\alpha - n^{1-s}\ln n, n\alpha]$ is $\leq (1-n^{-s}\ln
n)^m \leq e^{-\ln n} = n^{-1}$, and likewise for the interval
$[n\alpha, n\alpha + n^{1-s}\ln n]$, so the probability of ii) is
$\leq 2/n$.  Finally, if we are given that no bin is within
$n^{1/2}\ln n$ of $n\alpha$ (i.e. i) hasn't occured), then using a
Chernoff bound we can show that the probability of iii) is less than
$\exp(-\cO(\ln^2n))$.  Thus, the possibility of failure is dominated
by the probability of i), which is $\cO(n^{s-1/2}\ln n)$.

We say that the gentle measurement is successful if none of i), ii) or
iii) occur.  In this case, we can take as our estimate for $\alpha$ an
arbitrary value within the bin we have measured and by ii) will err by
no more than $2n^{-s}\ln n$.  Finally, let $M_j'$ be the measurement
outcome we obtain, let $\ket{\varphi}_{AB}$ be a purification of
$\rho_A^{\otimes n}$ and define $\pi:=M_j'\otimes I_B$.  Then the
post-measurement state is $\ket{\varphi'} =
\frac{\pi\ket{\varphi}}{\sqrt{\bra{\varphi}\pi\ket{\varphi}}}$ and
the entanglement fidelity is $F_e=\braket{\varphi}{\varphi'} =
\frac{\bra{\varphi}\pi\ket{\varphi}}
{\sqrt{\bra{\varphi}\pi\ket{\varphi}}} =
\sqrt{\bra{\varphi}\pi\ket{\varphi}}$.  From iii) we have
$\bra{\varphi}\pi\ket{\varphi}\geq 1-\epsilon$ where
$\epsilon=\exp(-cO(\ln^2n))$, so $F_e \geq \sqrt{1-\epsilon} = 1 -
\exp(-cO(\ln^2n))$.\footnote{A similar result was proved in
Lemma~9 of \cite{Wint99}}\qed

To perform gentle tomography we simply divide the $n$ states into
$d(d^2-1)$ blocks of length $l=\lfloor\frac{n}{d(d^2-1)}\rfloor$ and
gently measure each block.  If $\{\ket{v_i^{(k)}}\}_{i=1}^d$ is the
basis for $\sigma_k$, then we can index the blocks by $i=1,\ldots d$
and $k=1,\ldots,d^2-1$ and measure $\ket{v_i^{(k)}}$ on block
$(i,k)$.  

\begin{proposition}[Gentle tomography]
For any $0<s<1/2$ and fixed Hilbert space dimension, applying the
procedure described above to $\rho^{\otimes n}$ requires poly$(n)$
time and fails with probability $\cO(n^{s-1/2}\ln n)$.  Upon success,
the disturbance is less than $\cO(n^{-2})$ and the 
estimate $\tilde{\rho}$ satisfies $\|\rho-\tilde{\rho}\|_1\leq\cO(n^{-s}\ln
n)$.
\end{proposition}

{\bf Proof:} 
We say that tomography succeeds when each of the $d(d^2-1)$
measurements succeed individually.  Since the dimension $d$ is a
constant, we can use Proposition~\ref{prop:gentle} to bound the
failure probability by $\cO(d^{3(\frac{3}{2}-s)}n^{s-1/2}\ln n) \sim 
\cO(n^{s-1/2}\ln n)$ and the state disturbance by $\cO(d^9n^{-2}) \sim
\cO(n^{-2})$. 

We still need to describe how to form an accurate
estimate $\tilde{\rho}$.  Assume that each gentle measurement has
succeeded.  Then the $d(d^2-1)$ gentle measurements output not state
estimates, but bins, ($b_1,b_2,\ket{\phi}$), guaranteeing only that
$b_1\leq \bra{\phi}\rho\ket{\phi} \leq b_2$.  We will try to find a
state $\tilde{\rho}$ that is consistent with each bin.  Since $\rho$
is consistent with each bin, we know that some such $\tilde{\rho}$
exists.  We can find it efficiently by solving a semi-definite program
for $\tilde{\rho}$
given by the constraints: $\tilde{\rho}\geq 0$, $\tr\tilde{\rho}=1$
and $b_1\leq \bra{\phi}\tilde{\rho}\ket{\phi} \leq b_2$ for each bin
$(b_1,b_2,\ket{\phi})$.\footnote{ If one of the gentle
measurements fails, this semidefinite program may fail or it may
report a totally erroneous answer.}

Given such a $\tilde{\rho}$, we have for each gentle measurement that
$|\bra{\phi}(\rho-\tilde{\rho})\ket{\phi}|<\epsilon$, where
$\epsilon=\cO(n^{1-s}\ln n)$.  Then if $\sigma_k=\sum_i \lambda_i
\ket{v_i}\bra{v_i}$, $|\tr (\rho-\tilde{\rho})\sigma_k|=|\sum_i
\lambda_i \bra{v_i}(\rho-\tilde{\rho})\ket{v_i}| \leq \epsilon \sum_i
|\lambda_i| \leq \sqrt{d}\epsilon$.  Thus, by the Cauchy-Schwartz
inequality,
$$ \|\rho-\tilde{\rho}\|_1 \leq d \|\rho-\tilde{\rho}\|_2
 = d \sqrt{\sum_k (\tr (\rho-\tilde{\rho})\sigma_k)^2}
\leq d^{5/2}\epsilon$$
\qed

This extends our trade-off curve for gentle measurements to full
gentle state tomography.  It is an interesting question whether the
tradeoff we have found between accuracy and probability of failure is
optimal up to logarithmic factors.

\section{Universal compression}
Now look more closely at the quantum coding.  Schumacher compression
works by identifying the eigenvalues and eigenvectors of $\rho$, then
coherently performing classical Shannon compression on sequences of
those eigenvectors with probabilities given by the corresponding
eigenvalues.  However, we are forced to operate with only an estimate
$\tilde{\rho}\approx\rho$, so we will need to use a data compression
scheme that deals well with small inaccuracies in the state estimate.

This case has been analyzed in \cite{JP02}, which found that
compressing $\rho$ in the basis $\{\ket{i}\}$ with any classical
algorithm gives an asymptotic rate of $R=\sum_i \bra{i}\rho\ket{i}
\log \bra{i}\rho\ket{i}$.  This is because compressing $\rho$
faithfully reduces to compressing the diagonal entries of $\rho$ in an
arbitrary basis $\{\ket{i}\}$.  Due to the nonnegativity of the
relative entropy ($S(\rho\|\sigma)=\tr\rho(\log \rho-\log\sigma)\geq
0$), we have $R\leq-\tr\rho\log\sigma = S(\rho) + S(\rho\|\sigma)$ for
any density matrix $\sigma$ that can be diagonalized as $\sigma=\sum_i
p_i\ket{i}\bra{i}$.  Thus, for any density matrix $\sigma$, we can
encode $\rho$ by diagonalizing it in the basis of $\sigma$ and then
using a classical reversible algorithm.  This will achieve a rate $R
\leq S(\rho) + S(\rho\|\sigma)$.

Unfortunately, there is no simple bound for $S(\rho\|\tilde{\rho})$ in
terms of $\|\rho-\tilde{\rho}\|_1$; in fact, the relative entropy can
be infinite if the support of $\rho$ is not contained within the
support of $\tilde{\rho}$.  This problem corresponds to the situation
when our state estimate has led the encoder to believe that certain
vectors will never appear, so that when it encounters them in $\rho$,
it has made no provision to deal with them.  The solution to this is
simple: assume that any input vector has a small, but non-zero, chance
of occuring.  This means that instead of encoding according to
$\tilde{\rho}$, we will use $\tilde{\rho}_\delta:=(1-\delta)\tilde{\rho}+\delta
I/d$ as our state estimate, for some small $\delta>0$.

Suppose that after performing gentle tomography
$\|\tilde{\rho}-\rho\|_1<\epsilon$.  Then if we choose
$\epsilon,\delta = \cO(n^{-s}\log n)$, we can bound the rate by
\begin{eqnarray*}
R &\leq& -\tr\rho\log\tilde{\rho}_\delta \leq 
S(\tilde{\rho}_\delta)+\cO(n^{-s}\log^2 n)
\\&\leq& S(\rho) + \cO(n^{-s}\log^2 n)
\end{eqnarray*}
The second inequality follows from the operator inequality $\tilde{\rho}_\delta
\geq \delta I/d$ (implying $-\log\tilde{\rho}_\delta\geq\log (d/\delta) I =
\cO(\log n)I$) and the last inequality is due to Fannes' inequality.
We have neglected the inefficiency of the classical coding, since we
can choose it to be $\cO(n^{-s})$ and will incur only exponentially
small damage for $s<\frac{1}{2}$.

To analyze the errors, note that since we usually cannot tell when
tomography has failed, we ought to consider failure to be another form
of disturbance.  Thus, the $\cO(n^{s-\frac{1}{2}})$ probability of
failure dominates the state disturbance and the errors from classical
coding. This is consistent with the observation in \cite{HM02a} that
universal compression schemes have yet to achieve better than a
polynomially vanishing error.

Since our compression algorithm outputs a variable number of qubits,
damage to the encoded state is not the only possible form of error.
Upon failure, our algorithm risks producing a string length well above
the $n(S(\rho)+n^{-s}\log^2n)$ qubits we expect; in fact, the only absolute
bound we can establish is $n\log d$ qubits.  Fortunately, the
probability that $\rho^{\otimes n}$ is compressed to $nR$ qubits for
$R>S(\rho)$ decreases as $\cO(\exp(-nK))$ for some constant $K$
depending only on $\rho$ and $R$.  Following \cite{HM02a}, we define
this {\em overflow exponent} as
\be K = \lim_{n\rightarrow\infty} \frac{-1}{n}\log
\left[\mbox{prob. that $\rho^{\otimes n}$ yields
$\geq nR$ qubits}\right] \ee
The codes described in \cite{HM02a} achieve the
optimal value of $K$: $\inf_{\sigma:H(\sigma)\geq R}
S(\sigma\|\rho)$.  In contrast, 
 our algorithm\footnote{It is possible to gently measure
$\tr\rho\sigma_k$ directly, instead of inferring it from $d$ gentle
measurements of $\sigma_k$'s eigenvectors.  Using this for gentle
tomography results in a compression scheme with an overflow exponent
$d$ times higher, though still not optimal.}
 achieves
\be K=\inf_{\sigma:H(\sigma)\geq R}
\frac{1}{d(d^2-1)}\sum_{k=1}^{d^2-1} S(M_k(\sigma)\|M_k(\rho)) \ee
where $M_k$ denotes the operation of
measuring in the eigenbasis of $\sigma_k$ (i.e. $M_k(\rho)=\sum_i
\ket{v_i^{(k)}}\bra{v_i^{(k)}}\rho\ket{v_i^{(k)}}\bra{v_i^{(k)}}$).

To review, our encoding procedure is:
\begin{enumerate}
\item Perform gentle tomography on $\rho^{\otimes n}$ using
$n^s$ bins, yielding an estimate $\tilde{\rho}$.
\item Construct a modified estimate
$\tilde{\rho}_\delta=(1-\delta)\tilde{\rho}+\delta I/d$ for $\delta=\cO(n^{-s})$.
\item Encode $\rho^{\otimes n}$ with an efficient classical algorithm
(such as arithmetic coding\cite{CM00}) using the basis of
$\tilde{\rho}_\delta$ as the computational basis.
\item Attach a classical description of $\tilde{\rho}_\delta$ with $\cO(\sqrt{n})$
bits of precision and a $\lceil\log (n\log d)\rceil$ bit register
indicating the length of the compressed data.
\end{enumerate}
The decoding procedure is simply to extract the description of $\tilde{\rho}_\delta$
and use it as the basis for a classical decoding algorithm.

\section{Conclusion}
We have described a polynomial time algorithm for compressing
$\rho^{\otimes n}$ into $nS(\rho) + \cO(n^{-s}\log^2n)$ qubits with
error rate $\cO(n^{s-\frac{1}{2}}\log n)$.  This matches the error
rate and inefficiency of the proof of \cite{HM02a}, though not
their overflow exponent.  The procedure of \cite{JP02}, on the other
hand, can only achieve a compression rate of $S(\rho)+\cO(n^{-s})$ by
incurring an error rate of $\cO(n^{-\frac{1}{2} + s(1+d^2)})$ (possibly
up to logarithmic factors) and an overflow exponent of zero.  For
example, compressing qubits with constant error is only possible at a
rate of $S(\rho)+\cO(n^{-1/10})$.

More elegant would be a method for ergodic sources analogous to
Lempel-Ziv-Walsh coding 
that adaptively created a quantum dictionary and compressed quantum
information on the fly.  But the method proposed here still allows
the coding of sources with unknown statistics to attain the quantum
transmission limit for sources with known statistics as the message
length approaches infinity.

\section{Acknowledgements}
This work was partially supported by the Hewlett Packard--MIT
foundation (HP-MIT), by the ARO under a MURI program, by ARDA via NRO,
and by the NSA and ARDA under contract number DAAD19-01-1-06.  We are
grateful to I. Chuang, K. Matsumoto, R. Schack and B.
Schumacher for helpful discussions.


\begin{thebibliography}{99}
\newcommand{\bi}[1]{\bibitem{#1}}

\bi{S95} B. Schumacher, {\it Phys. Rev. A} {\bf 51}, 2738 (1995).
R. Jozsa and B. Schumacher, {\it J. Mod. Opt.}  {\bf 41} 2343 (1994).

\bi{JHHH98} R. Jozsa, M. Horodecki, P. Horodecki, R. Horodecki,
{\it Phys. Rev. Lett.} {\bf 81}, 1714 (1998).

\bi{CTDSZ80} C.M. Caves, K.S. Thorne, R.W.P. Drever, V.D.
Sandberg, M. Zimmerman, {\it Rev. Mod. Phys.} {\bf 52}, 341-392 (1980).

\bi{LS00} S. Lloyd and J.-J.E. Slotine, 
{\it Phys. Rev. A} {\bf 62}, 012307 (2000).

\bi{JP02} R. Jozsa and S. Presnell, arXiv:quant-ph/0210196 (2002).

\bi{HL04} A. Harrow and S. Lloyd, in preparation.

\bi{HM02a} M. Hayashi and K. Matsumoto, arXiv:quant-ph/0202001 (2002).  M. Hayashi and K. Matsumoto, arXiv:quant-ph/0209124 (2002).

\bi{HM02b} M. Hayashi and K. Matsumoto, arXiv:quant-ph/0209030 (2002).


\bi{BBPS95} C.H. Bennett, H.J. Bernstein, S. Popescu, B. Schumacher,
arXiv:quant-ph/9511030 (1995).

\bi{Wint99} A. Winter, {\it IEEE Trans. Inf. Theory}, {\bf 45}, 2481 
(1999).

\bi{CM00} I.L. Chuang, D.S. Modha, {\it IEEE Trans. Inf. Theory} {\bf
46}, 1104 (2000).

\end{thebibliography}
\end{document}